\documentclass[twoside]{article}
\usepackage{amsmath}
\usepackage[title]{appendix}
\usepackage[english]{babel}
\usepackage{float}
\usepackage{hyperref}

\newcommand{\ket}[1]{\left \vert #1 \right \rangle}

\usepackage{graphicx}
\usepackage[utf8]{inputenc}
\usepackage{nicefrac}
\usepackage{pgfplots}
\usepackage{placeins}
\usepackage{subcaption}
\usepackage{stmaryrd}
\usepackage{qic}
\usepackage{tikz}
\usetikzlibrary{shapes.geometric, calc}
\usepackage{verbatim}
\usepackage{xspace}

\tikzset{hilite/.style={line width=4pt, line cap=round, line join=round}}
\newcommand{\myGlobalTransformation}[2]
{
    \pgftransformcm{1}{0}{0.4}{0.5}{\pgfpoint{#1cm}{#2cm}}
}

\newcommand{\gridThreeD}[3]
{
    \begin{scope}
        \myGlobalTransformation{#1}{#2};
        \draw [#3,step=1.5cm] grid (4.49,4.49);
    \end{scope}
}

\tikzstyle myBG=[line width=3pt,opacity=1.0]

\providecommand \cfour {$\llbracket 4,\,2,\,2 \rrbracket$\xspace} 
\providecommand \name {\cfour-toric code\xspace} 
\providecommand{\abs}[1]{\left \vert #1 \right \vert}
\providecommand{\cnot}{\textsc{cnot}\xspace}
\providecommand{\cz}{\textsc{cz}\xspace}

\def \octagon {--++(0,2)--++(1,1)--++(2,0)--++(1,-1)--++(0,-2)--++(-1,-1)--++(-2,0)--cycle}
\def \btrap {--++(1,1)--++(2,0)--++(1,-1)--++(-4,0)--cycle}
\def \ltrap {--++(2,0)--++(1,1)--++(0,2)--++(-3,-3)--cycle}
\def \rtrap {--++(0,-2)--++(1,-1)--++(2,0)--++(-3,3)--cycle}
\definecolor{rd}{RGB}{200,0,0}
\definecolor{gn}{RGB}{0,200,0}
\definecolor{bl}{RGB}{0,0,200}
\newlength\figureheight
\newlength\figurewidth
\pgfplotsset{every tick label/.append style={font=\footnotesize}}
\pgfplotsset{every axis legend/.append style={font=\footnotesize}}
\pgfdeclarelayer{fg}    
\pgfsetlayers{main,fg}
\begin{document}
\setlength{\textheight}{8.0truein}    

\runninghead{Noise Thresholds for the  \cfour{}-concatenated Toric Code}
            {Criger/Terhal}

\normalsize\textlineskip
\thispagestyle{empty}
\setcounter{page}{1}

\copyrightheading{0}{0}{2003}{000--000}

\vspace*{0.88truein}

\alphfootnote

\fpage{1}

\centerline{\bf
\uppercase{Noise Thresholds for the  \cfour{}-concatenated Toric Code}}
\vspace*{0.37truein}
\centerline{\footnotesize
\textsc{Ben Criger}}
\vspace*{0.015truein}
\centerline{\footnotesize\it JARA Institute for Quantum Information, RWTH Aachen, Otto-Blumenthal-Stra{\ss}e 20}
\baselineskip=10pt
\centerline{\footnotesize\it Aachen, Nord-Rhein Westfalen, 52074 Germany}
\vspace*{10pt}
\centerline{\footnotesize 
\textsc{Barbara M. Terhal}}
\vspace*{0.015truein}
\centerline{\footnotesize\it JARA Institute for Quantum Information, RWTH Aachen, Otto-Blumenthal-Stra{\ss}e 20}
\baselineskip=10pt
\centerline{\footnotesize\it Aachen, Nord-Rhein Westfalen, 52074 Germany}
\vspace*{0.225truein}
\publisher{(received date)}{(revised date)}

\vspace*{0.21truein}

\abstracts{
We analyze the properties of a 2D topological code derived by concatenating the \cfour code with the toric/surface code, or alternatively by removing check operators from the 2D square-octagon or 4.8.8 color code.
We show that the resulting code has a circuit-based noise threshold of $\sim 0.41\%$ (compared to $\sim 0.6\%$ for the toric code in a similar scenario), which is higher than any known 2D color code.
We believe that the construction may be of interest for hardware in which one wants to use both long-range two-qubit gates as well as short-range gates between small clusters of qubits.
}{}{}

\vspace*{10pt}

\vspace*{3pt}
\communicate{to be filled by the Editorial}

\vspace*{1pt}\textlineskip    

\section{Introduction}
\label{sec:intro}

Quantum error correction (QEC) is believed to be a necessity for quantum computing. 
Two-dimensional topological quantum error correcting codes are the leading contenders for the implementation of quantum error correction due to the practical appeal of 2D qubit connectivity, relatively high noise thresholds, and universal fault-tolerant gate constructions (see e.g. the review \cite{Terhal:RMP}). 
Two popular families of 2D topological codes are \emph{toric/surface codes} \cite{KitaevToricCode} and \emph{color codes} \cite{bombin2006topological}, which we briefly introduce in Sections \ref{subsec:toric_surface_codes} and \ref{subsec:color_codes} for completeness.
The surface code architecture is based on planar qubit connectivity of low degree (each qubit participating in four parity checks, each parity check involving four qubits), and has a very high noise threshold $0.6\%-1\%$ \cite{Groszkowski, PhysRevA.83.020302}, when all gate error rates are identical. 
These advantages make it the focus of current experimental research. 

In this manuscript, we consider a scenario in which multi-qubit gates may have different error rates, with long-range gates having higher rates than short-range gates.
Previous work \cite{HierarchicalSurfaceCode} has considered extremely noisy long-range operations which connect relatively large regions of low-noise surface code tiles. 
We consider a scenario in which only a small region, called a cluster, is connected by short-range gates, with error rates for short- and long-range operations that are similar, though not necessarily identical. 

In order to arrive at a more clustered layout, we simply concatenate the toric code with a four-qubit code which encodes two qubits, the \cfour code. 
This code by itself is a natural testbed to assess the quality of parity check measurements as it is the smallest code which can detect a single error. 
Alternatively, the \cfour{} code can correct a single amplitude-damping error \cite{LeungNielsenChuangYamamoto,FletcherShorWin}, thus providing protection against $T_1$ errors. 
For these reasons implementing the \cfour{} code is a natural choice for early quantum error correction experiments.  
One can imagine that the four qubits in each cluster are coupled to ancilla qubits for parity check measurement using short-range gates.
Once the capacity of each four-qubit cluster to detect errors is established, one can tie these clusters together by surface code parity checks, concatenating the toric code with the \cfour{} code.
The inter-cluster toric code parity check measurements could be performed using long-range gates. 
Long-range connections could also be established by creating and distilling entangled states between clusters.
The point of this work is not to prescribe how to use such a \cfour-concatenated surface code (see Section \ref{sec:discussion} for ideas on superconducting qubit hardware, \cite{BrellNJP2011, TerhalHasslerDiV} for implementations of similar codes using perturbative gadgets and Majorana fermions), but to show that it may offer more flexibility in designing a physical layout while attaining a high noise threshold, similar to the surface code.

Coincidentally, the \cfour-concatenated toric/surface code (or \name{} for short) is identical to a reduced or `light' version of a 2D color code.
Two-dimensional color codes \cite{bombin2006topological} are 2D topological codes which have an advantage over the surface code in the transversality, and thus $O(1)$ time and space overhead, of the Hadamard and the $S$ gate (note that these gates are not transversal in the \name{}, see the discussion at the end of Section \ref{sec:construction}). It has also been argued in \cite{LandahlRyanAnderson} that,
despite the lower noise threshold, the color code qubit overhead is lower than that of the surface code. 

Another possible advantage of the \name{} over the surface code is that it can interface, using lattice surgery \cite{VT:lattice_surgery}, with a 3D color code in which the $T$ gate is implemented transversally \cite{bombin:transversalT}.
An encoded $T$ state ancilla can, through this method, be transferred fault-tolerantly to a planar array for further Clifford processing. 
The Clifford processing itself consists of \cnot gates which can be realized on qubits encoded in surface code or triangular color code sheets using lattice surgery along boundaries of the sheets \cite{horsman:suture,Terhal:RMP, LandahlRyanAnderson}. 
It may be possible to do a phase $S={\rm diag}(1,i)$ and Hadamard gate in the \name{} or a plain surface code if one encodes in lattice defects as in \cite{hastings_geller:dislocation}.

2D color codes have lower thresholds than the surface code, limiting their experimental feasibility in the near term.
This is often attributed to the increased weight of the check operators (six for a hexagonal color code, and four or eight for the square-octagon color code).
In the circuit-based noise model, which we also use in this paper (see Section \ref{sec:decoding}), one assumes that all elementary operations in the parity check circuits undergo errors.
For example, using a non-scalable decoder based on integer programming, a threshold of $0.082\%$ has been obtained for the triangular square-octagon (4.8.8) color code \cite{colorCodeThresholds}, although the noise threshold assuming noiseless parity check measurements was found to be almost identical to that of the toric code. 
Earlier work \cite{wang+:color} estimated the noise threshold of triangular color codes to be approximately $0.1\%$. 
More recent work \cite{stephens:threshold} has shown, using a decoder which maps the decoding problem onto three copies of a surface code decoding problem \cite{MultiMWPMDecoder}, that a circuit-based noise threshold for the 4.8.8 color code of $0.14\%$ can be attained. 
The high threshold of the 2D color codes for noiseless parity check measurements (also called the code-capacity threshold) suggests that their poor performance against circuit-based noise is due to the diminished reliability of the syndrome information.
In a circuit-based error model, the reliability of the syndrome is directly affected by the number of qubits involved in the parity check measurement as it determines the number of \cnot gates. 

In this paper, we show that a `light' version of the 2D color code, namely the \name{}, in which we measure fewer parity checks than in a standard color code, has quite a high noise threshold, almost the same as the toric code itself. 
This shows that the lower threshold of the color code is not solely due to measuring high-weight check operators. 
We arrive at these results by showing how the decoding problem is identical to that of the toric code and by carefully optimizing the parity check circuits.

\subsection{Overview}

In Sections \ref{subsec:toric_surface_codes} and \ref{subsec:color_codes} we review the toric/surface code and 2D color codes.
In Section \ref{sec:construction} we detail the check operators and logical operators of the \name{}. 
We show how it can be obtained from the square-octagon color code on a torus. 
In Section \ref{sec:decoding} we show that the \cfour-toric code can be decoded using minimum-weight perfect matching with a deformed Manhattan metric.
In Section \ref{sec:thresholds} we present the noise thresholds obtained against various error models. 
We conclude in Section \ref{sec:discussion} with a discussion.


\subsection{Toric/Surface Codes}
\label{subsec:toric_surface_codes}
The standard example of a 2D topological code is the toric code \cite{KitaevToricCode, DKLP}. 
It is defined by placing weight-four parity check operators on an $l$-by-$l$ square lattice as in Figure \ref{fig:toric_code}.
\begin{figure}
\centering
\begin{tikzpicture}
\draw[step=1cm] (0,0) grid (4.99,4.99);
\draw[black, line width=14pt, line cap=round, line join=round] (1, 1) rectangle (2, 2);
\draw[white, line width=12pt, line cap=round, line join=round] (1, 1) rectangle (2, 2);
\draw[black] (1,1.5) node{$Z$};
\draw[black] (1.5,2) node{$Z$};
\draw[black] (2,1.5) node{$Z$};
\draw[black] (1.5,1) node{$Z$};
\draw[black, line width=14pt, line cap=round, line join=round] (3, 2) -- (3, 4);
\draw[black, line width=14pt, line cap=round, line join=round] (2, 3) -- (4, 3);
\draw[white, line width=12pt, line cap=round, line join=round] (3, 2) -- (3, 4);
\draw[white, line width=12pt, line cap=round, line join=round] (2, 3) -- (4, 3);
\draw[black] (2.25,3) node{$X$};
\draw[black] (3.75,3) node{$X$};
\draw[black] (3,2.25) node{$X$};
\draw[black] (3,3.75) node{$X$};
\draw[black, line width=14pt, line cap=round, line join=round] (0, 0) -- (0, 5);
\draw[white, line width=12pt, line cap=round, line join=round] (0, 0) -- (0, 5);
\draw[black] (0,2.5) node {$\bar{Z}_1$};
\draw[black, line width=14pt, line cap=round, line join=round] (0, 0) -- (5, 0);
\draw[white, line width=12pt, line cap=round, line join=round] (0, 0) -- (5, 0);
\draw[black] (2.5,0) node {$\bar{Z}_2$};
\draw[black, line width=14pt, line cap=round, line join=round] (0, 4.5) -- (5, 4.5);
\draw[white, line width=12pt, line cap=round, line join=round] (0, 4.5) -- (5, 4.5);
\draw[black] (2.5,4.5) node {$\bar{X}_1$};
\draw[black, line width=14pt, line cap=round, line join=round] (4.5, 0) -- (4.5, 5);
\draw[white, line width=12pt, line cap=round, line join=round] (4.5, 0) -- (4.5, 5);
\draw[black] (4.5,2.5) node {$\bar{X}_2$};
\draw[black, line width=14pt, line cap=round, line join=round] (4, 4.5) -- (5, 4.5);
\draw[white, line width=12pt, line cap=round, line join=round] (3, 4.5) -- (5, 4.5);
\end{tikzpicture}
\vspace*{10pt}\fcaption{The toric code on a 5-by-5 lattice with periodic boundary conditions in both dimensions.
In this code a qubit is associated with each edge of the lattice, with stabilizer check operators associated with the vertices and faces.
The boundaries of faces (also called \emph{plaquettes}) support a check of the form $Z^{\otimes 4}$, the co-boundaries of vertices (also called \emph{stars}), support checks of the form $X^{\otimes 4}$.
Sets of edges that wrap around the torus support the two sets of logical $\overline{X}$ and $\overline{Z}$ operators, as shown above. }
\label{fig:toric_code}
\end{figure}
A toric code on an $l$-by-$l$ lattice has $2l^2$ physical qubits, two logical qubits and a distance $d=l$; it is a $\left \llbracket 2l^2,\, 2,\, l \right \rrbracket $ code. 
The surface code which encodes a single qubit can be obtained by imposing boundary conditions producing an $\left \llbracket l^2+(l-1)^2,\, 1,\, l \right \rrbracket $ code \cite{BK:surface}. 
A qubit-overhead optimized surface code encoding a single qubit can also be constructed with parameters $\left \llbracket l^2,\, 1,\, l \right \rrbracket$, see Figure~\ref{fig:rot_surface}.

\begin{figure}
\centering
\begin{tikzpicture}[x=0.75cm,y=0.75cm]
\foreach \x in {0.5, 2.5, 4.5}{
    \foreach \y in {0, 5}{
    \filldraw[draw=black, fill=white, line width=1 pt] (\x, \y) circle (0.5);        
    }
}
\foreach \y in {1.5, 3.5}{
    \foreach \x in {0, 5}{
    \filldraw[draw=black, fill=gray, line width=1 pt] (\x, \y) circle (0.5);        
    }
}
\foreach \x in {0, 2, 4}{
    \foreach \y in {0, 2, 4}{
        \fill[gray] (\x, \y) rectangle +(1,1);    
    }
}
\foreach \x in {1, 3}{
    \foreach \y in {1, 3}{
        \fill[gray] (\x, \y) rectangle +(1,1);    
    }
}
\foreach \x in {0, 4}{
    \foreach \y in {1, 3}{
        \fill[white] (\x, \y) rectangle +(1,1);    
    }
}
\draw[step=1, line width=1pt, line cap=round, line join=round] (0,0) grid (5,5);
\end{tikzpicture}
\vspace*{10pt}\fcaption{Surface code with qubits on vertices and parameters $\left \llbracket l^2,\, 1,\, l \right \rrbracket$, shown is $l=6$. 
Grey faces are $X$ checks and white faces are $Z$ checks. 
We can put a copy of this code with $X$ and $Z$ interchanged on top of this code lattice such that there are two qubits per vertex. If we encode those two qubits using the \cfour{} code, one obtains a 4.8.8 color code encoding two logical qubits.}
\label{fig:rot_surface}
\end{figure}

\subsection{Color Codes}
\label{subsec:color_codes}
The two-dimensional color codes are obtained by placing qubits on the vertices of a lattice, with both an $X$ and $Z$ stabilizer check $S_u(X)$ (resp. $S_u(Z)$) on each face $u$ of the lattice \cite{bombin2006topological}.
The lattice must be \emph{trivalent} (all vertices having degree three) and face three-colorable (the set of faces must be divisible into three subsets such that no face from a given subset is adjacent to another face from the same subset).
This guarantees that each face is supported on an even number of qubits, and that adjacent faces share exactly two vertices \cite{KubicaBeverland}, implying that all check operators commute. Examples of lattices obeying these trivalent and three-colorable constraints are a lattice with hexagonal plaquettes and the square-octagon lattice (see Figure \ref{fig:squoct_lattice}).
\begin{figure}
\centering
\begin{tikzpicture}[scale=0.4]
\clip[draw] (0,0) rectangle (10,10);
\foreach \x in {0, 3, 6, 9}{
    \foreach \y in {-1, 2, 5, 8}{
        \filldraw[fill=rd, draw=black] (\x, \y) rectangle ++(2,2);    
    }
}

\foreach \x in {-1, 5}{
    \foreach \y in {-1, 5}{
        \filldraw[fill=bl,draw=black] (\x, \y) \octagon;    
    }
}
\foreach \x in {2, 8}{
    \foreach \y in {2, 8}{
        \filldraw[fill=gn, draw=black] (\x, \y) \octagon;    
    }
}
\end{tikzpicture}
\vspace*{10pt}\fcaption{A section of a square-octagon lattice.
Each vertex has degree three, and the faces are divided into three subsets, such that no two faces from the same subset are adjacent. 
This lattice can, therefore, be used to define a 2D color code.
\label{fig:squoct_lattice}
}
\end{figure}
Note that one can associate a color to each edge so that the edge joins two plaquettes of that color.
For a lattice on the torus there is a linear dependency between the check operators, namely the product of all $Z$ stabilizer (resp. $X$) checks of one color equals the product of $Z$ (resp. $X$) check operators of another color, i.e. $\Pi_{u \in {\rm Blue}} S_u(X)=\Pi_{u \in {\rm Green}} S_u(X)=\Pi_{u \in {\rm Red}} S_u(X)$ and $\Pi_{u \in {\rm Blue}} S_u(Z)=\Pi_{u \in {\rm Green}} S_u(Z)=\Pi_{u \in {\rm Red}} S_u(Z)$.
In order to calculate the number of logical qubits which are encoded by the lattice, one can use the expression for the Euler characteristic, $\chi = \abs{V} - \abs{E} + \abs{F}$ ($\chi = 0$ for the torus). 
Here $V$ is the set of vertices, $E$ the set of edges and $F$ the set of faces of the lattice.
Trivalence of the lattice implies that $\abs{E}=\frac{3}{2} \abs{V}$ so that $\abs{V}=2\abs{F}$; the total number of qubits is even. 
The number of logical qubits for the torus is the difference between the number of physical qubits and the number of linearly-independent checks, in this case $\abs{V} - 2\abs{F} + 4 = 4$.
The logical operators $\overline{X}_i,\overline{Z}_i, i=1,\ldots, 4$ form non-trivial loops, running over edges of a specific color, around the torus, similar to the toric code. 

One can also define a color code with open boundaries, for example a triangular color code, see Figure~\ref{fig:triangle}.
One can obtain this triangular code by puncturing a trivalent lattice which covers the sphere.
Puncturing means that one removes a qubit and all the check operators which act on it.
For the sphere one has $\abs{V} - \abs{E} + \abs{F} = 2$ so that $\abs{V}=2\left( \abs{F} - 2 \right)$, again even.
The unpunctured code then encodes no qubits.
When one removes a single qubit (and its associated checks, a total of six), the total number of physical qubits is odd.
The triangular code encodes a single qubit and the oddness of the number of physical qubits implies that $\overline{X}=X_{\rm all}$ and $\overline{Z}=Z_{\rm all}$ form a pair of anti-commuting logical operators ($X_{\rm all}$ acts as Pauli $X$ on all qubits in the lattice). A multi-qubit color code can be obtained using a polygon of higher degree \cite{KYP:unfolding}.
By multiplying these logical operators with check operators, one can deform the logical operators to operators on a boundary of the triangular lattice.
One can associate a color to each boundary; the color of the removed check operator.

For this lattice, the parameters of the triangular code are $\left \llbracket n=\frac{1}{2}d^2+d-\frac{1}{2},\, 1,\, d \right \rrbracket $ \cite{colorCodeThresholds}.
2D triangular color codes allow a transversal logical Hadamard and phase gate, that is, $\overline{H}=H^{\otimes n}$, and $\overline{S} = S^{\otimes T} \otimes S^{\dagger \otimes \overline{T}}$ with $T$ being a subset of the vertices \cite{KubicaBeverland}. In addition, the logical \cnot can be performed using lattice surgery \cite{horsman:suture,Terhal:RMP,LandahlRyanAnderson}.

\begin{figure}
\centering
\mbox{
\begin{tikzpicture}[x=0.4cm, y=0.4cm]
\filldraw[draw=black, fill=gn, line cap=round, line join=round] (0,0) \ltrap;
\filldraw[draw=black, fill=gn, line cap=round, line join=round] (6,6) \ltrap;
\foreach \x in {2, 8, 14}{
\filldraw[draw=black, fill=bl, line cap=round, line join=round] (\x,0) \btrap;
}
\foreach \x/\y in {3/1, 6/4, 9/7, 9/1, 12/4, 15/1}{
\filldraw[draw=black, fill=rd, line cap=round, line join=round] (\x,\y) rectangle ++(2,2);
}
\filldraw[draw=black, fill=gn, line cap=round, line join=round] (5,1) \octagon;
\filldraw[draw=black, fill=gn, line cap=round, line join=round] (11,1) \octagon;
\filldraw[draw=black, fill=bl, line cap=round, line join=round] (8,4) \octagon;
\filldraw[draw=black, fill=bl, line cap=round, line join=round] (14,6) \rtrap;
\end{tikzpicture}
}
\vspace*{10pt}\fcaption{Triangular color code encoding one qubit with $d = 7$: the logical string operators (either $\overline{X}$ or $\overline{Z}$) can run along the three boundaries and commute with all checks.
}
\label{fig:triangle}
\end{figure}

\section{The \name{}}
\label{sec:construction}
We introduce a topological code on a square-octagon lattice with qubits on its vertices.
We can create this code by concatenating the toric code (or a surface code) with the $\llbracket 4,\,2,\,2 \rrbracket$ code \cite{KnillC4}.
The stabilizer group of the $\llbracket 4,\,2,\,2 \rrbracket$ code is generated by the check operators $XXXX$ and $ZZZZ$ and the code encodes two qubits with logical operators: 
\begin{equation}
\overline{X}_1 = XIXI,\,\overline{X}_2 = IIXX,\,\overline{Z}_1 = IZIZ,\,\overline{Z}_2 = ZZII. \nonumber
\end{equation}
Placing the qubits on the corners of a square results in a layout for the check and logical operators seen in Figure \ref{fig:422}.
\begin{figure}
\centering
\resizebox{\textwidth}{!}{
\begin{tikzpicture}
\draw[hilite] (0, 0) rectangle +(2, 2);
\draw[black, line width=14 pt, line cap=round, line join=round] (0.375,0.375) rectangle +(1.25,1.25);
\draw[white, line width=12 pt, line cap=round, line join=round] (0.375,0.375) rectangle +(1.25,1.25);
\draw (1.625,1.625) node {$X$};
\draw (0.375,1.625) node {$X$};
\draw (1.625,0.375) node {$X$};
\draw (0.375,0.375) node {$X$};
\draw[hilite] (3, 0) rectangle +(2, 2);
\draw[black, line width=14 pt, line cap=round, line join=round] (3.375,0.375) rectangle +(1.25,1.25);
\draw[white, line width=12 pt, line cap=round, line join=round] (3.375,0.375) rectangle +(1.25,1.25);
\draw (4.625,1.625) node {$Z$};
\draw (3.375,1.625) node {$Z$};
\draw (4.625,0.375) node {$Z$};
\draw (3.375,0.375) node {$Z$};
\draw[hilite] (6, 0) rectangle +(2, 2);
\draw[black, line width=14 pt, line cap=round, line join=round] (6.375,1.625) -- +(1.25,0);
\draw[white, line width=12 pt, line cap=round, line join=round] (6.375,1.625) -- +(1.25,0);
\draw (7.625,1.625) node {$X$};
\draw (6.375,1.625) node {$X$};
\draw[hilite] (9, 0) rectangle +(2, 2);
\draw[black, line width=14 pt, line cap=round, line join=round] (9.375,0.375) -- +(0, 1.25);
\draw[white, line width=12 pt, line cap=round, line join=round] (9.375,0.375) -- +(0, 1.25);
\draw (9.375,1.625) node {$X$};
\draw (9.375,0.375) node {$X$};
\draw[hilite] (12, 0) rectangle +(2, 2);
\draw[black, line width=14 pt, line cap=round, line join=round] (13.625,0.375) -- +(0, 1.25);
\draw[white, line width=12 pt, line cap=round, line join=round] (13.625,0.375) -- +(0, 1.25);
\draw (13.625,0.375) node {$Z$};
\draw (13.625,1.625) node {$Z$};
\draw[hilite] (15, 0) rectangle +(2, 2);
\draw[black, line width=14 pt, line cap=round, line join=round] (15.375,0.375) -- +(1.25,0);
\draw[white, line width=12 pt, line cap=round, line join=round] (15.375,0.375) -- +(1.25,0);
\draw (16.625,0.375) node {$Z$};
\draw (15.375,0.375) node {$Z$};
\end{tikzpicture}
}
\vspace*{10pt}\fcaption{The two stabilizer checks and four logical operators for the $\llbracket 4,\,2,\,2 \rrbracket$ code, placed on the vertices of a square.}
\label{fig:422}
\end{figure}

To concatenate, we replace every qubit of the toric code with a square or cluster with the \cfour code defined on its vertices:
\begin{figure}[H]
\centering
\resizebox{0.5\textwidth}{!}{
\begin{tikzpicture}
\draw[hilite] (0,0) -- (6,0);
\draw[hilite] (0,0) -- (0,6);
\draw[->, hilite] (8, 3) -- (10,3);
\draw[hilite] (11, 2) rectangle + (2,2);
\draw[hilite] (14, -1) rectangle + (2,2);
\draw[hilite] (12,0) -- (18,0);
\draw[hilite] (12,0) -- (12,6);
\end{tikzpicture}
}
\vspace*{10pt}\fcaption{Each edge on the original toric code lattice is replaced with a square containing four qubits. }
\end{figure}

To produce the concatenated code check operators, the check operators of the toric code are written in terms of the logical operators of the $\llbracket 4,\,2,\,2 \rrbracket$ code, and the check operators of each cluster are added.
We have the option of choosing one of the logical qubits of the $\llbracket 4,\,2,\,2 \rrbracket$ code to be a \emph{gauge} qubit, i.e. a qubit that will not be used and whose state can be left to vary or be fixed.
We will select the logical qubit such that the logical $Z$ (to become part of a $\overline{Z}^{\otimes 4}$ check) is on a side of the square which is parallel to the edge of the underlying square lattice.
With this selection, the toric code check operators become the check operators seen in Figure \ref{fig:422toric} when written using the logical operators of the $\llbracket 4,\,2,\,2 \rrbracket$ code.
\begin{figure}
\centering
\begin{tikzpicture}[x=0.25cm,y=0.25cm]
\foreach \x in {0, 6, 12, 18}{
    \foreach \y in {0, 6, 12, 18}{
        \draw[black!30, line width=2pt, line join=round, line cap=round] (\x, \y) -- +(0, 6);
        \draw[black!30, line width=2pt, line join=round, line cap=round] (\x, \y) -- +(6, 0);
        \draw (\x-1, \y+2) rectangle +(2,2);
        \draw (\x+2, \y-1) rectangle +(2,2);
        \draw (\x+1, \y+4) -- +(1, 1);
        \draw (\x+4, \y+1) -- +(1, 1);
        \draw (\x+1, \y+2) -- +(1, -1);   
        \draw (\x+4, \y-1) -- +(1, -1);   
    }
}
\draw[black, line width = 6pt, line join=round] (7, 8) \octagon;
\draw[bl, line width = 4pt, line join=round] (7, 8) \octagon;
\draw[black, line width = 6pt, line join=round] (16, 11) \octagon;
\draw[gn, line width = 4pt, line join=round] (16, 11) \octagon;
\draw[black, line width = 6pt, line join=round] (5, 14) rectangle +(2,2);
\draw[rd, line width = 4pt, line join=round] (5, 14) rectangle +(2,2);
\draw[black, line width = 14pt, line cap=round] (-1, 2) -- + (20,0);
\draw[white, line width = 12pt, line cap=round] (-1, 2) -- + (20,0);
\draw (9,2) node {$\bar{X}_1$};
\draw[black, line width = 14pt, line cap=round] (22, -1) -- + (0, 20);
\draw[white, line width = 12pt, line cap=round] (22, -1) -- + (0, 20);
\draw (22,9) node {$\bar{X}_2$};
\draw[black, line width = 14pt, line cap=round] (2, 19) -- + (20, 0);
\draw[white, line width = 12pt, line cap=round] (2, 19) -- + (20, 0);
\draw (11,19) node {$\bar{Z}_2$};
\draw[black, line width = 14pt, line cap=round] (-1, 2) -- + (0, 20);
\draw[white, line width = 12pt, line cap=round] (-1, 2) -- + (0, 20);
\draw (-1, 11) node {$\bar{Z}_1$};
\end{tikzpicture}
\vspace*{10pt}\fcaption{Check operators and logical Pauli operators for the \name .
The blue octagon supports a $Z$ check, the green octagon supports an $X$ check, the red squares support {\rm both} $X$ and $Z$ checks.}
\label{fig:422toric}
\end{figure}

This concatenated code, the \name{}, encodes $2$ logical qubits (the logical qubits of the toric code) and $2l^2$ gauge qubits (one gauge qubit per cluster of four qubits).
The total number of qubits is $8 l^2$ and the distance of the code is $2l$ as the logical operator of the toric code must have support in $l$ separate clusters and on two qubits per cluster.
Thus the parameters of this code are $\llbracket 8l^2,\,2,\,2l \rrbracket$, showing that the qubit overhead for a given distance is identical to the toric code.

A `full' color code on the 4.8.8 lattice can also be obtained by concatenating the \cfour{} code and the toric code.
This requires \emph{two} toric codes for concatenation, one with $Z$ checks on plaquettes and $X$ checks on stars as usual, and one with $Z$ checks on stars and $X$ checks on plaquettes. 
This construction can also use two rotated surface codes (see Figure~\ref{fig:rot_surface}), resulting in a color code on a lattice with four open boundaries, encoding two logical qubits. 

General relations showing that $D$-dimensional color codes can be viewed as multiple copies of surface codes up to local unitary transformations and addition and removal of ancilla qubits exist, even in the presence of boundaries, see e.g. \cite{KYP:unfolding,BDP:uni,Bombin2014}.
Thus, we see a very concrete realization of this mapping using code concatenation: the code concatenation perspective directly suggests ways of doing noisy syndrome measurements and implementing a decoder.
One can obtain the 4.6.12 color code lattice \cite{colorCodeThresholds} from an identical concatenation step of two copies of the toric code with the \cfour{} code: one starts with a hexagonal toric code lattice with qubits on edges. 
The code has hexagonal weight-six plaquettes and weight-three stars which upon concatenation with the \cfour{} code become weight-twelve checks and weight-six checks respectively. 
Similarly, it is possible to obtain the 6.6.6 color code from concatenating the toric code and the $\llbracket 6,4,2\rrbracket$ code, see Appendix \ref{app:642}. 


The difference between the full 4.8.8 color code and the \name{} lies in the fact that the full color code encodes an additional logical qubit whose logical operators are string operators on the gauge qubits of \name{}. 
In the \name{}, we do not correct for errors on these gauge qubits, nor do we fix their states by check measurements. 
In other words, a simpler construction for the \name{ }begins with a color code defined on a square-octagonal tiling and removes $X$ checks from blue octagons and $Z$ checks from green octagons. 
In doing so, $XX$ edges between green $X$-octagons become undetectable, as do $ZZ$ edges between blue $Z$ octagons.
These operators are precisely the gauge operators of the \cfour code defined earlier, as seen in Figure \ref{fig:gauge_ops}.
\begin{figure}
\centering
\begin{tikzpicture}[x=0.5cm, y=0.5cm]
\filldraw[fill=rd] (0,0) rectangle +(2,2) ;
\filldraw[fill=gn] (-4,0) \octagon;
\filldraw[fill=gn] (2,0) \octagon;
\filldraw[fill=bl] (-1,-3) \octagon;
\filldraw[fill=bl] (-1,3) \octagon;
\draw[black, line width=14pt, line cap=round] (0.25, -0.5) -- (1.75, -0.5);
\draw[white, line width=12pt, line cap=round] (0.25, -0.5) -- (1.75, -0.5);
\draw (0.25,-0.5) node {$X$};
\draw (1.75,-0.5) node {$X$};
\draw[black, line width=14pt, line cap=round] (0.25, 2.5) -- (1.75, 2.5);
\draw[white, line width=12pt, line cap=round] (0.25, 2.5) -- (1.75, 2.5);
\draw (0.25,2.5) node {$X$};
\draw (1.75,2.5) node {$X$};
\draw[black, line width=14pt, line cap=round] (-0.5, 0.25) -- (-0.5, 1.75);
\draw[white, line width=12pt, line cap=round] (-0.5, 0.25) -- (-0.5, 1.75);
\draw (-0.5, 0.25) node {$Z$};
\draw (-0.5, 1.75) node {$Z$};
\draw[black, line width=14pt, line cap=round] (2.5, 0.25) -- (2.5, 1.75);
\draw[white, line width=12pt, line cap=round] (2.5, 0.25) -- (2.5, 1.75);
\draw (2.5, 0.25) node {$Z$};
\draw (2.5, 1.75) node {$Z$};
\end{tikzpicture}
\vspace*{10pt}\fcaption{Gauge logical operators produced by removing stabilizer checks from octagonal faces as defined above.
These operators are precisely the logical operators for the unused gauge qubits in the \cfour { }code used in the concatenation scheme.}
\label{fig:gauge_ops}
\end{figure}

Given this relation between the surface code and the 4.8.8 color code, one can ask about the properties of the triangular version of the concatenated surface code.
This code is defined as the triangular color code, with $X$ and $Z$ checks removed from blue and green plaquettes, respectively.
Unfortunately, one can now multiply the logical operator of the encoded qubit by the weight-two gauge operators to reduce its weight to one, see Figure~\ref{fig:badtriangle}.
To see this, take a logical $\overline{Z}$ running along the green boundary in Figure~\ref{fig:triangle}: on each square touching this boundary we can remove its support by multiplication with a blue $ZZ$ edge.
Thus, for the qubit encoded in the triangular code to have high distance, one needs to fix (and thus measure) all the stabilizer checks of the color code.
This is unfortunate, as the triangular color code has a transversal $S$ gate, which arises partially due to the oddness of the number of physical qubits.
It is obvious that any concatenation of a surface code or copies of a surface code with the \cfour{} code always results in a logical $\overline{X}$ which acts on an even number of qubits, thus simple transversality of the $S$ gate is excluded for any \cfour{}-concatenated surface code.


\begin{figure}
\centering
\begin{tikzpicture}[x=0.5cm, y=0.5cm]
\filldraw[draw=black, fill=gn, line cap=round, line join=round] (0,0) \btrap;
\filldraw[draw=black, fill=gn, line cap=round, line join=round] (6,0) \btrap;
\filldraw[draw=black, fill=gn, line cap=round, line join=round] (1,3) \ltrap;
\filldraw[draw=black, fill=bl, line cap=round, line join=round] (9,3) \rtrap;
\filldraw[draw=black, fill=bl, line cap=round, line join=round] (3,1) \octagon;
\filldraw[draw=black, fill=rd, line cap=round, line join=round] (1,1) rectangle ++(2,2);
\filldraw[draw=black, fill=rd, line cap=round, line join=round] (4,4) rectangle ++(2,2);
\filldraw[draw=black, fill=rd, line cap=round, line join=round] (7,1) rectangle ++(2,2);
\draw[black, line width=14pt, line cap=round] (0,0)--(0,0);
\draw[white, line width=12pt, line cap=round] (0,0)  node[black]{$X$} -- (0,0);
\draw[black, line width=14pt, line cap=round] (12,0)--(12,0);
\draw[white, line width=12pt, line cap=round] (12,0)  node[black]{$Z$} -- (12,0);
\end{tikzpicture}
\vspace*{10pt}\fcaption{Triangular boundary conditions for the \name induce low-weight logical operators which are shown.}
\label{fig:badtriangle}
\end{figure}
In the following section, we compare the performance of the toric code and the \name{}, focusing on the error threshold.
\section{Decoding}
\label{sec:decoding}
To determine the ability of topological codes to correct errors, they are commonly subjected to random errors using three scenarios:
\begin{description}
\item[Data-Only Errors] in which every qubit in the lattice is acted upon by a Pauli $X$ and $Z$ with probability $p$.
These errors are detected at the points where they anticommute with local stabilizer checks.
\item[Data \& Syndrome Errors] in which errors act on the lattice as above, and syndrome measurements are subject to symmetric bit-flip with probability $q$ (usually set to $p$ for simplicity).
In this scenario, faulty measurements are repeated, typically $d$ times.
The resulting measurement record is then decoded.
\item[Circuit-Based Errors] in which errors act on the lattice and syndrome measurements as above, but faulty operations are modeled by perfect operations, followed by a random Pauli, sampled uniformly from the set of one- or two-qubit Paulis \cite{Groszkowski}. 
In this work, we assume that each operation (including the identity) is subject to the same error rate $p$, unless otherwise noted. 
\end{description}
In each of these scenarios, a logical error rate can be estimated by performing a large number of Monte Carlo trials of the appropriate decoding algorithm ($10^4$ for Figures \ref{fig:data_only}, \ref{fig:data_syndrome}, \ref{fig:first_noisy}, \ref{fig:full_monty}, and \ref{fig:oct_factor_3} below).

The pattern of syndromes generated by errors acting on the color code is different from that of the toric code.
Instead of indicating the endpoints of connected chains, the violated stabilizer checks may also be at the endpoints of Y-shaped \emph{string-nets} (see Figure \ref{fig:string_nets}).
\begin{figure}
\centering
\begin{tikzpicture}[scale=0.20]
\def \octagon {--++(0,2)--++(1,1)--++(2,0)--++(1,-1)--++(0,-2)--++(-1,-1)--++(-2,0)--cycle}
\clip[draw] (0,0) rectangle (20,20);
\foreach \x in {0, 3, 6, 9, 12, 15, 18}{
    \foreach \y in {-1, 2, 5, 8, 11, 14, 17}{
        \filldraw[fill=rd, draw=black] (\x, \y) rectangle ++(2,2);    
    }
}

\foreach \x in {-1, 5, 11, 17}{
    \foreach \y in {-1, 5, 11, 17}{
        \filldraw[fill=bl,draw=black] (\x, \y) \octagon;    
    }
}
\foreach \x in {2, 8, 14, 20}{
    \foreach \y in {2, 8, 14, 20}{
        \filldraw[fill=gn, draw=black] (\x, \y) \octagon;    
    }
}
\filldraw[draw=white,fill=black!66] (6,16) circle (0.3 cm);
\filldraw[draw=white,fill=black!66] (8,16) circle (0.3 cm);
\filldraw[draw=white,fill=black!66] (4,15) circle (0.5cm);
\filldraw[draw=white,fill=black!66] (10,15) circle (0.5cm);
\filldraw[draw=white,fill=black!33] (15,7) circle (0.3 cm);
\filldraw[draw=white,fill=black!33] (14,10) circle (0.3 cm);
\filldraw[draw=white,fill=black!33] (12,8) circle (0.3 cm);
\filldraw[draw=white,fill=black!33] (13,12) circle (0.5 cm);
\filldraw[draw=white,fill=black!33] (10,9) circle (0.5 cm);
\filldraw[draw=white,fill=black!33] (16,6) circle (0.5 cm);
\end{tikzpicture}
\vspace*{10pt}\fcaption{The syndrome pattern produced by $X$ or $Z$ errors on the highlighted vertices, for error chains which produce two or three violated stabilizer checks.}
\label{fig:string_nets}
\end{figure}
This additional complication in the structure of the syndromes makes the problem of finding optimal decoders for color codes more difficult. 
There are decoders based on the renormalization group \cite{RGDecoder}, and decoders which decompose the decoding problem into multiple instances of minimum-weight perfect matching \cite{MultiMWPMDecoder}. 
Using a non-scalable decoder based on integer programming, thresholds of $10.56\%$ (data-only errors), $3.05\%$ (data $\&$ syndrome errors) and $0.082\%$ (circuit-based errors) have been obtained for the color code on the square-octagon lattice \cite{colorCodeThresholds}. 
In \cite{andrist+:color} the highest possible noise threshold against data \& syndrome errors occurring with equal probability was found to be $4.5\%$ for the 6.6.6 2D color code.

For the toric code the decoding problem is well understood. 
To restore the code state, it is necessary to assign an error to the observed syndrome such that, with high probability, the product of the assigned error with the actual error is an element in the stabilizer group. 
A generated pattern of syndromes does not correspond uniquely to the error which caused it, because any connected chain of errors on data qubits or syndrome bits produces syndromes at its endpoints (see Figure \ref{fig:syndrome_pairs}).
\begin{figure}
\centering
\begin{minipage}{5cm}
\begin{tikzpicture}
\draw[step=1cm] (0,0) grid (4.99,4.99);
\draw[black!33!white, hilite] (1,2) -- (1,3) -- (2,3) -- (3,3);
\draw[black!33!white, hilite, loosely dashed] (1,2) -- (2,2) -- (3,2) -- (3,3);
\draw[black!66!white, hilite] (1,2) -- (0,2);
\draw[black!66!white, hilite] (5,2) -- (4,2) -- (4,3) -- (3,3);
\fill[black] (1, 2) circle(0.25cm);
\fill[black] (3, 3) circle(0.25cm);
\end{tikzpicture}
\end{minipage}
\hspace{1cm}
\begin{minipage}{5cm}
\begin{tikzpicture}
\gridThreeD{0}{0}{black};
\begin{scope}
    \myGlobalTransformation{0}{0};
    \foreach \x in {0, 1, 2} {
        \foreach \y in {2, 1, 0} {
            \node (star\x;\y) at (1.5*\x, 1.5*\y) {};
            \node (plaquette\x;\y) at (1.5*\x+0.75,1.5*\y+0.75) {};
            {
                \pgftransformreset
                \foreach \z in {0, 1, 2}{
                    \draw[black] (star\x;\y) -- ++(0,\z);
                    \draw[black] (plaquette\x;\y) -- ++(0,\z);
                }
                \foreach \z in {0, 1, 2}{
                \filldraw[fill=white] (star\x;\y) + (0,\z) circle (0.1cm);
                \filldraw[fill=white] (plaquette\x;\y) + (0,\z) circle (0.1cm);                
                }
            }
        }
    }
    \node (synd_one) at (1.5, 1.5) {};
    \node (synd_two) at (3, 3) {};
    \node (corner_one) at (3, 1.5) {};
    \node (corner_two) at (3, 3) {};
	\draw[color=black!50, line width=0.2cm, line cap=round, line join=round] (1.5,1.5) -- (3,1.5) -- (3, 3);    
    \pgftransformreset
    \filldraw[fill=black] (synd_one) circle (0.1cm);
    \draw[color=black!50, line width=0.2cm, line cap=round, line join=round] (synd_two) -- ++ (0,1);
    \filldraw[fill=black] (synd_two) + (0,1) circle (0.1cm);
    \draw[black] (star2;0) --+ (0,0.875);
    \draw[black] (plaquette1;0) --+ (0,0.875);
    \draw[black] (plaquette2;0) -- ++(0,2);
    \foreach \z in {0, 1, 2}{
        \filldraw[fill=white] (plaquette2;0) + (0,\z) circle (0.1cm);
    }
\end{scope}
\end{tikzpicture}
\end{minipage}
\vspace*{10pt}\fcaption{Left: a syndrome in two dimensions, which can be generated by $Z$ errors occurring on the highlighted edges.
The light gray paths are equivalent up to the action of an element of the stabilizer group.
The dark gray path differs from the light gray paths by the action of a logical operator. 
Right: a syndrome in three dimensions, which can be generated by $Z$ errors occurring on the highlighted horizontal edges and measurement errors occurring on the vertical highlighted edges. }
\label{fig:syndrome_pairs}
\end{figure}

Thus, decoding for the toric code can be accomplished by noting that the negative log-likelihood of a pattern of errors forming a continuous chain is proportional to a weighted Manhattan distance between the endpoints of the chain \cite{DKLP}.
The problem is then reduced to finding a set of pairs which minimizes the total length of the assigned error chains; a well-studied combinatorial problem called \emph{minimum-weight perfect matching} \cite{Edmonds}.
Under the three error models described above, using an MWPM decoder, the toric code has threshold error parameters of $10.3\%$, $2.9\%$ and $0.6\%$, respectively \cite{Groszkowski, HarringtonThesis}, identical to our results for the toric code in Figures~\ref{fig:data_only}, \ref{fig:data_syndrome} and \ref{fig:first_noisy}.\\

Let us now examine the decoding problem for the \name{}: it is also solvable by minimum-weight perfect matching of a set of points in 2D or 3D.
As in the toric and color codes, the CSS nature of the \name{} permits $X$ and $Z$ syndromes to be decoded as though they are caused by uncorrelated errors, with a small loss in threshold if errors are in fact correlated (though decoders exist which can compensate for this \cite{CorrelatedDecoder}).
As in the toric code, these syndromes occur at the endpoints of error chains.
Figure~\ref{fig:chains} shows some examples of error chains and their endpoints, which can be a pair of squares, a pair of octagons or a square-octagon pair. 
Again, we see that we only need to find a minimum-length path between non-trivial syndromes and paths which are the same in length (touch the same number of qubits) are equivalent given the gauge qubit degrees of freedom.
\begin{figure}
\centering
\begin{tikzpicture}[x=10pt, y=10pt]
\draw[black!30, step=6, line width=2pt, line cap=round, line join=round] (0,0) grid (17.99,17.99);
\foreach \x in {0, 6, 12}{
    \foreach \y in {0, 6, 12}{
        \draw[line width=2pt] (\x+2, \y-1) rectangle +(2,2);
        \draw[line width=2pt] (\x-1, \y+2) rectangle +(2,2);
        \draw[line width=2pt] (\x+1, \y+2) \octagon;
    }
}
\draw[black, line width=14pt, line cap=round, line join=round] (4,1) -- ++(1,1) -- ++(0,2) -- ++(0,4) -- ++(0,2)  ;
\draw[white, line width=12pt, line cap=round, line join=round] (4,1)node[black]{$Z$} -- ++(1,1)node[black]{$Z$} -- ++(0,2)node[black]{$Z$} -- ++(0,4)node[black]{$Z$} -- ++(0,2) node[black]{$Z$} ;
\filldraw[draw=white,fill=black!66] (3,0) circle (7pt);
\filldraw[draw=white,fill=black!66] (6,12) circle (7pt);

\filldraw[draw=white,fill=black!66] (9,15) circle (7pt);
\filldraw[draw=white,fill=black!66] (15,3) circle (7pt);

\draw[black, line width=14pt, line cap=round, line join=round] (10,13) -- ++(0,-2) -- ++(0,-4) -- ++(0,-2) -- ++(1,-1) -- ++ (2, 0)  ;
\draw[white, line width=12pt, line cap=round, line join=round] (10,13)  node[black]{$X$} -- ++(0,-2) node[black]{$X$} -- ++(0,-4) node[black]{$X$} -- ++(0,-2) node[black]{$X$} -- ++(1,-1) node[black]{$X$} -- ++ (2, 0) node[black]{$X$} ;

\filldraw[draw=white,fill=black!66] (3,12) circle (7pt);
\filldraw[draw=white,fill=black!66] (3,15) circle (7pt);

\draw[black, line width=14pt, line cap=round, line join=round] (2,13) -- ++(0,0);
\draw[white, line width=12pt, line cap=round, line join=round] (2,13)  node[black]{$X$} -- ++(0,0);

\draw[black, line width=14pt, line cap=round, line join=round] (14,11) -- ++(0,2);
\draw[white, line width=12pt, line cap=round, line join=round] (14,11)  node[black]{$Z$} -- ++(0,2) node[black]{$Z$};

\filldraw[draw=white,fill=black!66] (0,6) circle (7pt);
\filldraw[draw=white,fill=black!66] (0,12) circle (7pt);
\draw[black, line width=14pt, line cap=round, line join=round] (1,8) -- ++(0,2);
\draw[white, line width=12pt, line cap=round, line join=round] (1,8)  node[black]{$Z$} -- ++(0,2) node[black]{$Z$};
\end{tikzpicture}
\vspace*{10pt}\fcaption{Examples of $Z$ and $X$ error chains ending at defects (stabilizer checks with a syndrome indicating the error type).
Consider a single $X$ error which is detected by the adjacent square and octagon, see top-left.
An $X$ error on the qubit immediately to the right would have caused the same syndrome, but we do not need to distinguish between the two errors as their product is a gauge qubit logical $\overline{X}=XX$.
Two adjacent $Z$ errors on a square can either be a gauge qubit $\overline{Z}$, going undetected (on the right), or detected by two adjacent octagons (on the left).
In the latter case, three other $ZZ$ errors on the same square could have caused the same syndrome, but all are related by stabilizer $ZZZZ$ check and gauge qubit $\overline{Z}=ZZ$ on the square, hence we do not need to resolve this ambiguity.}
\label{fig:chains}
\end{figure}
We can thus consider two sublattices, one for correction of $Z$ and one for the correction of $X$ errors: the sublattices are formed by $X$-squares and $X$-octagons and $Z$-squares and $Z$-octagons resp., see one sublattice in Figure~\ref{fig:sublattice}.

\begin{figure}
\centering
\begin{tikzpicture}[x=10pt,y=10pt]
\foreach \x in {0, 6}{
    \foreach \y in {0, 6}{
        \filldraw[fill=black!30, line width=2pt, line cap=round, line join=round] (\x-1, \y+2) rectangle +(2,2);
        \filldraw[fill=black!30, line width=2pt, line cap=round, line join=round] (\x+2, \y-1) rectangle +(2,2);
        \draw[line width=2pt, line cap=round, line join=round] (\x, \y) -- +(0, 6);
        \draw[line width=2pt, line cap=round, line join=round] (\x, \y) -- +(6, 0);
        \draw[line width=2pt, line cap=round, line join=round] (\x+1, \y+4) -- +(1, 1);
        \draw[line width=2pt, line cap=round, line join=round] (\x+4, \y+1) -- +(1, 1);
        \draw[line width=2pt, line cap=round, line join=round] (\x+1, \y+2) -- +(1, -1);   
        \draw[line width=2pt, line cap=round, line join=round] (\x+4, \y-1) -- +(1, -1);   
    }
}
\filldraw[fill=black!30, line width=2pt, line cap=round, line join=round] (1,2) \octagon ;
\filldraw[fill=black!30, line width=2pt, line cap=round, line join=round] (7,2) \octagon ;
\filldraw[fill=black!30, line width=2pt, line cap=round, line join=round] (1,8) \octagon ;
\filldraw[fill=black!30, line width=2pt, line cap=round, line join=round] (7,8) \octagon ;
\end{tikzpicture}
\vspace*{10pt}\fcaption{Square-octagon lattice, with faces shaded over which a path may be constructed to assign a length to a pair of syndromes indicating $X$ errors. }
\label{fig:sublattice}
\end{figure}
On the sublattice, one marks the checks with non-trivial syndromes (putting defect points on the dual lattice) and one constructs a matching of all defects such that it minimizes the total distance (over the lattice) between them.
For a pair of defects on octagons, the minimum path length can still be calculated using the Manhattan distance, taking into account toric boundary conditions as for the toric code. 
For defect pairs where one or both defects are on a square, the Manhattan distance may be smaller than the minimum path length, corresponding to a path which traverses a forbidden octagon. 
However, the correct distance will be given by the Manhattan-metric path connecting octagons neighbouring the square endpoints, with an additional unit of length per square endpoint.
Each square has two nearest-neighbour octagons per syndrome type, so at most four octagon-octagon distances must be calculated per syndrome pair. 
This adds minimal overhead to decoding. 

For scenarios in which the stabilizer check measurements may themselves be in error, we consider the process of assigning a distance to changes in the measurement record which are separated in time and space.
In principle, we would assign different edge weights to time-like edges representing errors in the measurement of different check operators.
In order to provide an accurate comparison with the toric code, however, we use uniform edge weights to obtain thresholds in both the toric and \name{}s in the following section. 

We note that the mapping onto a minimum-weight matching problem (first used in \cite{hoelzer:thesis}) is the same as for the full color code decoder in \cite{MultiMWPMDecoder, stephens:threshold} where matchings are sought for three different sublattices, i.e. one sublattice of red squares and blue octagons, one sublattice of red squares and green octagons and one sublattice of green and blue octagons.
We have thus shown that due to the gauge qubit degree of freedom the reduced color code can be decoded by solving only one minimum-weight perfect matching problem.



\section{Thresholds}
\label{sec:thresholds}
We begin by comparing the \name{ }to the toric code in the ``Data-Only Errors'' scenario in Figure \ref{fig:data_only}.
\begin{figure}
\centering
\begin{tikzpicture}
\draw[draw=none, use as bounding box](0,0) rectangle (2.1\figurewidth,1.1\figureheight);
\definecolor{color1}{rgb}{0.880043634700096,0.147850027791703,0.349655219977259}
\definecolor{color0}{rgb}{0.917647058823529,0.917647058823529,0.949019607843137}
\definecolor{color3}{rgb}{0.143489950732319,0.522679613653864,0.359145694379759}
\definecolor{color2}{rgb}{0.514644709195696,0.464088173636572,0.137913981118522}
\definecolor{color5}{rgb}{0.71741510578539,0.191245761439708,0.905682494363787}
\definecolor{color4}{rgb}{0.157855715549606,0.500777210942904,0.586518795245686}

\begin{axis}[
title={Toric Code, Data-Only Errors},
xlabel={Physical Error Probability},
ylabel={Logical Error Probability},
xmin=0.08, xmax=0.12,
ymin=0, ymax=1,
width=\figurewidth,
height=\figureheight,
xtick={0.08,0.085,0.09,0.095,0.1,0.105,0.11,0.115,0.12},
xticklabels={0.080,0.085,0.090,0.095,0.100,0.105,0.110,0.115,0.120},
ytick={0,0.2,0.4,0.6,0.8,1},
yticklabels={0.0,0.2,0.4,0.6,0.8,1.0},
tick align=outside,
xmajorgrids,
x grid style={white},
ymajorgrids,
y grid style={white},
axis line style={white},
axis background/.style={fill=color0},
legend entries={{$d=8$},{$d=12$},{$d=16$},{$d=20$},{$d=24$}},
legend style={at={(0.025,0.95)},anchor=north west}
]
\addplot [draw=black, fill=color1, mark=*, only marks] table {%
0.08 0.238452309538
0.0821052631579 0.256948610278
0.0842105263158 0.281543691262
0.0863157894737 0.298140371926
0.0884210526316 0.324435112977
0.0905263157895 0.350129974005
0.0926315789474 0.370925814837
0.0947368421053 0.400419916017
0.0968421052632 0.414817036593
0.0989473684211 0.442111577684
0.101052631579 0.469306138772
0.103157894737 0.488702259548
0.105263157895 0.509398120376
0.107368421053 0.535792841432
0.109473684211 0.541891621676
0.111578947368 0.578784243151
0.113684210526 0.593781243751
0.115789473684 0.608578284343
0.117894736842 0.635272945411
0.12 0.651069786043
};
\addplot [draw=black, fill=color2, mark=*, only marks] table {%
0.08 0.173860911271
0.0821052631579 0.194044764189
0.0842105263158 0.21962430056
0.0863157894737 0.240707434053
0.0884210526316 0.277777777778
0.0905263157895 0.306354916067
0.0926315789474 0.329936051159
0.0947368421053 0.366207034373
0.0968421052632 0.388289368505
0.0989473684211 0.411171063149
0.101052631579 0.45243804956
0.103157894737 0.481414868106
0.105263157895 0.513988808953
0.107368421053 0.540667466027
0.109473684211 0.56554756195
0.111578947368 0.594024780176
0.113684210526 0.622402078337
0.115789473684 0.645883293365
0.117894736842 0.676758593125
0.12 0.693445243805
};
\addplot [draw=black, fill=color3, mark=*, only marks] table {%
0.08 0.1291
0.0821052631579 0.1468
0.0842105263158 0.1731
0.0863157894737 0.2006
0.0884210526316 0.2315
0.0905263157895 0.2641
0.0926315789474 0.3023
0.0947368421053 0.331
0.0968421052632 0.3727
0.0989473684211 0.4079
0.101052631579 0.4395
0.103157894737 0.4851
0.105263157895 0.52
0.107368421053 0.5482
0.109473684211 0.5903
0.111578947368 0.616
0.113684210526 0.6412
0.115789473684 0.684
0.117894736842 0.701
0.12 0.7285
};
\addplot [draw=black, fill=color4, mark=*, only marks] table {%
0.08 0.0886567164179
0.0821052631579 0.112736318408
0.0842105263158 0.143781094527
0.0863157894737 0.171343283582
0.0884210526316 0.201393034826
0.0905263157895 0.238109452736
0.0926315789474 0.272736318408
0.0947368421053 0.316517412935
0.0968421052632 0.344378109453
0.0989473684211 0.397512437811
0.101052631579 0.439701492537
0.103157894737 0.474129353234
0.105263157895 0.515422885572
0.107368421053 0.560099502488
0.109473684211 0.597313432836
0.111578947368 0.625970149254
0.113684210526 0.655422885572
0.115789473684 0.704577114428
0.117894736842 0.728656716418
0.12 0.751343283582
};
\addplot [draw=black, fill=color5, mark=*, only marks] table {%
0.08 0.0639800995025
0.0821052631579 0.0858706467662
0.0842105263158 0.108855721393
0.0863157894737 0.137114427861
0.0884210526316 0.166666666667
0.0905263157895 0.205870646766
0.0926315789474 0.242985074627
0.0947368421053 0.283482587065
0.0968421052632 0.331343283582
0.0989473684211 0.382288557214
0.101052631579 0.43184079602
0.103157894737 0.48447761194
0.105263157895 0.524577114428
0.107368421053 0.568756218905
0.109473684211 0.619701492537
0.111578947368 0.656119402985
0.113684210526 0.684676616915
0.115789473684 0.722885572139
0.117894736842 0.75144278607
0.12 0.784577114428
};
\path [draw=white, fill opacity=0] (axis cs:0.08,1)
--(axis cs:0.12,1);

\path [draw=white, fill opacity=0] (axis cs:1,0)
--(axis cs:1,1);

\path [draw=white, fill opacity=0] (axis cs:0.08,0)
--(axis cs:0.12,0);

\path [draw=white, fill opacity=0] (axis cs:0,0)
--(axis cs:0,1);

\end{axis}


\definecolor{color1}{rgb}{0.880043634700096,0.147850027791703,0.349655219977259}
\definecolor{color0}{rgb}{0.917647058823529,0.917647058823529,0.949019607843137}
\definecolor{color3}{rgb}{0.143489950732319,0.522679613653864,0.359145694379759}
\definecolor{color2}{rgb}{0.514644709195696,0.464088173636572,0.137913981118522}
\definecolor{color5}{rgb}{0.71741510578539,0.191245761439708,0.905682494363787}
\definecolor{color4}{rgb}{0.157855715549606,0.500777210942904,0.586518795245686}

\begin{axis}[
at={(1.1\figurewidth,0)},
title={\name{}, Data-Only Errors},
xlabel={Physical Error Probability},
ylabel={Logical Error Probability},
xmin=0.08, xmax=0.12,
ymin=0, ymax=1,
width=\figurewidth,
height=\figureheight,
xtick={0.08,0.085,0.09,0.095,0.1,0.105,0.11,0.115,0.12},
xticklabels={0.080,0.085,0.090,0.095,0.100,0.105,0.110,0.115,0.120},
ytick={0,0.2,0.4,0.6,0.8,1},
yticklabels={0.0,0.2,0.4,0.6,0.8,1.0},
tick align=outside,
xmajorgrids,
x grid style={white},
ymajorgrids,
y grid style={white},
axis line style={white},
axis background/.style={fill=color0},
legend entries={{$d=8$},{$d=12$},{$d=16$},{$d=20$},{$d=24$}},
legend style={at={(0.025,0.95)},anchor=north west}
]
\addplot [draw=black, fill=color1, mark=*, only marks] table {%
0.08 0.2795
0.0821052631579 0.2952
0.0842105263158 0.3158
0.0863157894737 0.3358
0.0884210526316 0.3614
0.0905263157895 0.3871
0.0926315789474 0.413
0.0947368421053 0.4372
0.0968421052632 0.4515
0.0989473684211 0.4767
0.101052631579 0.488
0.103157894737 0.5247
0.105263157895 0.5376
0.107368421053 0.5542
0.109473684211 0.5815
0.111578947368 0.598
0.113684210526 0.611
0.115789473684 0.6343
0.117894736842 0.6456
0.12 0.6689
};
\addplot [draw=black, fill=color2, mark=*, only marks] table {%
0.08 0.20375924815
0.0821052631579 0.233453309338
0.0842105263158 0.254849030194
0.0863157894737 0.28424315137
0.0884210526316 0.307338532294
0.0905263157895 0.350729854029
0.0926315789474 0.370525894821
0.0947368421053 0.397220555889
0.0968421052632 0.422415516897
0.0989473684211 0.453909218156
0.101052631579 0.47800439912
0.103157894737 0.517196560688
0.105263157895 0.536192761448
0.107368421053 0.567886422715
0.109473684211 0.581683663267
0.111578947368 0.621775644871
0.113684210526 0.646370725855
0.115789473684 0.659368126375
0.117894736842 0.684563087383
0.12 0.704959008198
};
\addplot [draw=black, fill=color3, mark=*, only marks] table {%
0.08 0.15307753797
0.0821052631579 0.187949640288
0.0842105263158 0.207833733014
0.0863157894737 0.234212629896
0.0884210526316 0.273181454836
0.0905263157895 0.303457234213
0.0926315789474 0.342825739408
0.0947368421053 0.375899280576
0.0968421052632 0.415367705835
0.0989473684211 0.441746602718
0.101052631579 0.492505995204
0.103157894737 0.51568745004
0.105263157895 0.549160671463
0.107368421053 0.57364108713
0.109473684211 0.607414068745
0.111578947368 0.641486810552
0.113684210526 0.66726618705
0.115789473684 0.689748201439
0.117894736842 0.709632294165
0.12 0.749700239808
};
\addplot [draw=black, fill=color4, mark=*, only marks] table {%
0.08 0.121593291405
0.0821052631579 0.145153239493
0.0842105263158 0.18219027653
0.0863157894737 0.201257861635
0.0884210526316 0.238195068384
0.0905263157895 0.275032444844
0.0926315789474 0.314665069382
0.0947368421053 0.352400918439
0.0968421052632 0.391734052111
0.0989473684211 0.432963961266
0.101052631579 0.47289607667
0.103157894737 0.521513427174
0.105263157895 0.559349106519
0.107368421053 0.598083258461
0.109473684211 0.634022162324
0.111578947368 0.661275831087
0.113684210526 0.690925426774
0.115789473684 0.723270440252
0.117894736842 0.745033443147
0.12 0.761705101328
};
\addplot [draw=black, fill=color5, mark=*, only marks] table {%
0.08 0.0928912783751
0.0821052631579 0.118279569892
0.0842105263158 0.14247311828
0.0863157894737 0.1729390681
0.0884210526316 0.208084428515
0.0905263157895 0.250099561928
0.0926315789474 0.296794105934
0.0947368421053 0.333333333333
0.0968421052632 0.38162086818
0.0989473684211 0.428016726404
0.101052631579 0.48476702509
0.103157894737 0.519912385504
0.105263157895 0.569394663481
0.107368421053 0.613201911589
0.109473684211 0.652528872959
0.111578947368 0.685284747113
0.113684210526 0.706093189964
0.115789473684 0.740143369176
0.117894736842 0.771903624054
0.12 0.789924332935
};
\path [draw=white, fill opacity=0] (axis cs:0.08,1)
--(axis cs:0.12,1);

\path [draw=white, fill opacity=0] (axis cs:1,0)
--(axis cs:1,1);

\path [draw=white, fill opacity=0] (axis cs:0.08,0)
--(axis cs:0.12,0);

\path [draw=white, fill opacity=0] (axis cs:0,0)
--(axis cs:0,1);

\end{axis}
\end{tikzpicture}
\vspace*{20pt}\fcaption{Threshold comparison between the toric code and the \name{}, for the ``Data-Only Errors'' model from Section \ref{sec:decoding}.}
\label{fig:data_only}
\end{figure}
Given that check operator weight has no effect on the data-only error model, it is perhaps expected that both possess a threshold near 10.3\%.
Nevertheless, this similarity serves to demonstrate the efficacy of the punctured metric introduced in Section \ref{sec:decoding}. 
We consider the ``Data \& Syndrome Errors'' scenario in Figure \ref{fig:data_syndrome}.
\begin{figure}
\begin{tikzpicture}
\draw[draw=none, use as bounding box](0,0) rectangle (2.1\figurewidth,1.1\figureheight);
\definecolor{color1}{rgb}{0.880043634700096,0.147850027791703,0.349655219977259}
\definecolor{color0}{rgb}{0.917647058823529,0.917647058823529,0.949019607843137}
\definecolor{color3}{rgb}{0.143489950732319,0.522679613653864,0.359145694379759}
\definecolor{color2}{rgb}{0.514644709195696,0.464088173636572,0.137913981118522}
\definecolor{color5}{rgb}{0.71741510578539,0.191245761439708,0.905682494363787}
\definecolor{color4}{rgb}{0.157855715549606,0.500777210942904,0.586518795245686}

\begin{axis}[
title={Toric Code, Data \& Syndrome Errors},
xlabel={Physical Error Probability},
ylabel={Logical Error Probability},
xmin=0.02, xmax=0.04,
ymin=0, ymax=1,
width=\figurewidth,
height=\figureheight,
ytick={0,0.2,0.4,0.6,0.8,1},
yticklabels={0.0,0.2,0.4,0.6,0.8,1.0},
tick align=outside,
xmajorgrids,
x grid style={white},
ymajorgrids,
y grid style={white},
axis line style={white},
axis background/.style={fill=color0},
legend style={at={(0.025,0.95)}, anchor=north west},
legend entries={{$d=8$},{$d=10$},{$d=12$},{$d=14$},{$d=16$}}
]
\addplot [draw=black, fill=color1, mark=*, only marks] table {%
0.02 0.01885
0.0210526315789 0.0245
0.0221052631579 0.031875
0.0231578947368 0.04375
0.0242105263158 0.05225
0.0252631578947 0.0691
0.0263157894737 0.085475
0.0273684210526 0.1076
0.0284210526316 0.130975
0.0294736842105 0.15695
0.0305263157895 0.18775
0.0315789473684 0.220275
0.0326315789474 0.25725
0.0336842105263 0.29775
0.0347368421053 0.3349
0.0357894736842 0.3773
0.0368421052632 0.420125
0.0378947368421 0.460475
0.0389473684211 0.50255
0.04 0.547025
};
\addplot [draw=black, fill=color2, mark=*, only marks] table {%
0.02 0.008775
0.0210526315789 0.012875
0.0221052631579 0.0205
0.0231578947368 0.02945
0.0242105263158 0.039825
0.0252631578947 0.053
0.0263157894737 0.0722
0.0273684210526 0.094
0.0284210526316 0.125625
0.0294736842105 0.155125
0.0305263157895 0.193675
0.0315789473684 0.238725
0.0326315789474 0.287325
0.0336842105263 0.33595
0.0347368421053 0.39275
0.0357894736842 0.447075
0.0368421052632 0.501625
0.0378947368421 0.5582
0.0389473684211 0.6072
0.04 0.65185
};
\addplot [draw=black, fill=color3, mark=*, only marks] table {%
0.02 0.003875
0.0210526315789 0.007275
0.0221052631579 0.011575
0.0231578947368 0.01875
0.0242105263158 0.028325
0.0252631578947 0.0404
0.0263157894737 0.06145
0.0273684210526 0.08445
0.0284210526316 0.11565
0.0294736842105 0.156175
0.0305263157895 0.2003
0.0315789473684 0.25695
0.0326315789474 0.32465
0.0336842105263 0.381125
0.0347368421053 0.455725
0.0357894736842 0.522375
0.0368421052632 0.591125
0.0378947368421 0.6479
0.0389473684211 0.70415
0.04 0.753
};
\addplot [draw=black, fill=color4, mark=*, only marks] table {%
0.02 0.002425
0.0210526315789 0.004125
0.0221052631579 0.0076
0.0231578947368 0.012825
0.0242105263158 0.01995
0.0252631578947 0.03215
0.0263157894737 0.051575
0.0273684210526 0.077425
0.0284210526316 0.111725
0.0294736842105 0.159675
0.0305263157895 0.211525
0.0315789473684 0.28385
0.0326315789474 0.356125
0.0336842105263 0.439175
0.0347368421053 0.518075
0.0357894736842 0.595375
0.0368421052632 0.670125
0.0378947368421 0.735125
0.0389473684211 0.784675
0.04 0.826125
};
\addplot [draw=black, fill=color5, mark=*, only marks] table {%
0.02 0.001025
0.0210526315789 0.0023
0.0221052631579 0.00445
0.0231578947368 0.007875
0.0242105263158 0.0146
0.0252631578947 0.0244
0.0263157894737 0.0415
0.0273684210526 0.067525
0.0284210526316 0.10255
0.0294736842105 0.1571
0.0305263157895 0.2231
0.0315789473684 0.3034
0.0326315789474 0.3933
0.0336842105263 0.490725
0.0347368421053 0.58745
0.0357894736842 0.665825
0.0368421052632 0.742825
0.0378947368421 0.7986
0.0389473684211 0.84255
0.04 0.874925
};
\path [draw=white, fill opacity=0] (axis cs:0.02,1)
--(axis cs:0.04,1);

\path [draw=white, fill opacity=0] (axis cs:1,0)
--(axis cs:1,1);

\path [draw=white, fill opacity=0] (axis cs:0.02,0)
--(axis cs:0.04,0);

\path [draw=white, fill opacity=0] (axis cs:0,0)
--(axis cs:0,1);

\end{axis}

\definecolor{color1}{rgb}{0.880043634700096,0.147850027791703,0.349655219977259}
\definecolor{color0}{rgb}{0.917647058823529,0.917647058823529,0.949019607843137}
\definecolor{color3}{rgb}{0.143489950732319,0.522679613653864,0.359145694379759}
\definecolor{color2}{rgb}{0.514644709195696,0.464088173636572,0.137913981118522}
\definecolor{color5}{rgb}{0.71741510578539,0.191245761439708,0.905682494363787}
\definecolor{color4}{rgb}{0.157855715549606,0.500777210942904,0.586518795245686}

\begin{axis}[
at={(1.1\figurewidth,0)},
title={\name{}, Data \& Syndrome Errors},
xlabel={Physical Error Probability},
ylabel={Logical Error Probability},
xmin=0.02, xmax=0.04,
ymin=0, ymax=1,
width=\figurewidth,
height=\figureheight,
ytick={0,0.2,0.4,0.6,0.8,1},
yticklabels={0.0,0.2,0.4,0.6,0.8,1.0},
tick align=outside,
xmajorgrids,
x grid style={white},
ymajorgrids,
y grid style={white},
axis line style={white},
axis background/.style={fill=color0},
legend style={at={(0.025,0.95)}, anchor=north west},
legend entries={{$d=8$},{$d=10$},{$d=12$},{$d=14$},{$d=16$}}
]
\addplot [draw=black, fill=color1, mark=*, only marks] table {%
0.02 0.0353
0.0210526315789 0.0423
0.0221052631579 0.054
0.0231578947368 0.0663
0.0242105263158 0.0786
0.0252631578947 0.102
0.0263157894737 0.1247
0.0273684210526 0.1516
0.0284210526316 0.1807
0.0294736842105 0.2124
0.0305263157895 0.2434
0.0315789473684 0.2758
0.0326315789474 0.3222
0.0336842105263 0.3644
0.0347368421053 0.4028
0.0357894736842 0.4529
0.0368421052632 0.4896
0.0378947368421 0.531
0.0389473684211 0.5645
0.04 0.6062
};
\addplot [draw=black, fill=color2, mark=*, only marks] table {%
0.02 0.0168
0.0210526315789 0.0259
0.0221052631579 0.0348
0.0231578947368 0.0428
0.0242105263158 0.0597
0.0252631578947 0.0757
0.0263157894737 0.099
0.0273684210526 0.1287
0.0284210526316 0.1691
0.0294736842105 0.2018
0.0305263157895 0.2398
0.0315789473684 0.2849
0.0326315789474 0.3405
0.0336842105263 0.3978
0.0347368421053 0.4428
0.0357894736842 0.497
0.0368421052632 0.5548
0.0378947368421 0.6088
0.0389473684211 0.6618
0.04 0.696
};
\addplot [draw=black, fill=color3, mark=*, only marks] table {%
0.02 0.0095
0.0210526315789 0.0117
0.0221052631579 0.0193
0.0231578947368 0.0289
0.0242105263158 0.0405
0.0252631578947 0.0554
0.0263157894737 0.0776
0.0273684210526 0.1102
0.0284210526316 0.1441
0.0294736842105 0.1941
0.0305263157895 0.2422
0.0315789473684 0.3054
0.0326315789474 0.3619
0.0336842105263 0.4368
0.0347368421053 0.4955
0.0357894736842 0.5574
0.0368421052632 0.6256
0.0378947368421 0.6807
0.0389473684211 0.7306
0.04 0.7743
};
\addplot [draw=black, fill=color4, mark=*, only marks] table {%
0.02 0.0053
0.0210526315789 0.0066
0.0221052631579 0.0114
0.0231578947368 0.0185
0.0242105263158 0.0305
0.0252631578947 0.0443
0.0263157894737 0.068
0.0273684210526 0.0959
0.0284210526316 0.1401
0.0294736842105 0.1855
0.0305263157895 0.2533
0.0315789473684 0.3151
0.0326315789474 0.3939
0.0336842105263 0.4837
0.0347368421053 0.5465
0.0357894736842 0.6354
0.0368421052632 0.693
0.0378947368421 0.7548
0.0389473684211 0.8021
0.04 0.8366
};
\addplot [draw=black, fill=color5, mark=*, only marks] table {%
0.02 0.0021
0.0210526315789 0.0038
0.0221052631579 0.0057
0.0231578947368 0.0127
0.0242105263158 0.0205
0.0252631578947 0.0317
0.0263157894737 0.0569
0.0273684210526 0.091
0.0284210526316 0.1224
0.0294736842105 0.179
0.0305263157895 0.2521
0.0315789473684 0.3366
0.0326315789474 0.4247
0.0336842105263 0.5138
0.0347368421053 0.6062
0.0357894736842 0.6858
0.0368421052632 0.7516
0.0378947368421 0.8114
0.0389473684211 0.8482
0.04 0.8793
};
\path [draw=white, fill opacity=0] (axis cs:0.02,1)
--(axis cs:0.04,1);

\path [draw=white, fill opacity=0] (axis cs:1,0)
--(axis cs:1,1);

\path [draw=white, fill opacity=0] (axis cs:0.02,0)
--(axis cs:0.04,0);

\path [draw=white, fill opacity=0] (axis cs:0,0)
--(axis cs:0,1);

\end{axis}

\end{tikzpicture}
\vspace*{20pt}\fcaption{Threshold comparison for an error model which includes syndrome errors with probability $p$ identical to the bit/phase-flip error rate.}
\label{fig:data_syndrome}
\end{figure}
Given that the addition to the error model is identical in both cases, it is not surprising to see that the threshold is the same, about $3\%$. 

A more detailed treatment of the circuit-based error model is necessary, in part to show the dependence of the error threshold on syndrome extraction circuit design. 
We treat the circuit-based model in the following subsection. 
\subsection{Circuit-Based Errors}
If we consider circuit-based errors, the form of the parity check circuits is relevant.
For the \name, one could in principle do quantum error correction with the typical procedure for concatenated codes.
This means that one encodes the parity check circuits for the toric code with the \cfour code (since the data qubits are also encoded with the \cfour code).
This means that a single ancilla qubit in $\ket{+}$ in a parity check circuit for the toric code gets replaced by four qubits encoded in $\ket{\overline{+}}$ of the \cfour{} code.
This four-qubit ancilla could also be verified so that any single error in the ancilla preparation can only lead to a single error on the four outgoing qubits (modulo gauge qubit errors).
A single \cnot gate in the toric code parity check circuit would be replaced by two \cnot gates. 
Between each \cnot gate one can possibly insert a quantum error correction step for the \cfour{} code, a measurement of $XXXX$ and $ZZZZ$, to detect whether errors have occurred.
We have chosen not to consider this fully concatenated parity check circuitry here as it requires more overhead, and the efficacy of this scheme is not immediately clear.
The question to be solved is how to optimally use the additional error detection information in the toric code decoder.
It is not excluded that the threshold of this kind of circuitry is {\em higher} than that of the toric code itself; after all the use of the \cfour{} code in concatenated form has been shown to numerically lead to a threshold of $3\%$ by Knill in \cite{KnillC4}.

First we assume that we extract the syndrome of all checks (octagon and square) using single-qubit ancillas and one round of quantum error correction consists of measuring all $Z$ and $X$ checks.
One can optimize the circuits used for syndrome extraction such that they take a minimal number of timesteps.
It is also possible to constrain the circuit to produce weight-two gauge qubit errors (of $X$ or $Z$ type) on clusters of four qubits.
For the toric code, interleaving leads to a single round of quantum error correction taking four timesteps (not counting preparation and measurement of the ancilla qubit) \cite{fowler+:practical}. We obtain a circuit which contains eight \cnot timesteps, shown in Figure \ref{fig:8_step_sched}.
\begin{figure}
\centering
\includegraphics[width=0.7\textwidth]{squoct_patch_two_rows.pdf}
\vspace*{10pt}\fcaption{\cnot schedule for syndrome extraction (shown are eight steps in which \cnot{}s occur). Not shown are initial state preparation ($\ket{0}$ for $Z$ checks and $\ket{+}$ for $X$ checks) and measurement (in the $X$ basis for $X$ checks and the $Z$ basis for $Z$ checks).
The ancillas associated with the weight-four check operators whose \cnot schedules are contained in the top row are measured (in one additional timestep) before the \cnot gates in the bottom row are executed (the remaining qubits can undergo errors during this single step measurement).
The square $X$ checks are measured first, followed by a measurement of the square $Z$ checks.}
\label{fig:8_step_sched}
\end{figure}
This increases the total probability of an error occurring during measurement, which is reflected in the lower threshold for the \name seen in Figure \ref{fig:first_noisy}.
\begin{figure}
\centering

\vspace*{20pt}\fcaption{Circuit-based noise threshold comparison using the parity check circuits in Figure~\ref{fig:8_step_sched}. Error bars are $95\%$ confidence intervals.}
\label{fig:first_noisy}
\end{figure}

This threshold can be increased, however, by noting that the gate schedule in Figure~\ref{fig:8_step_sched} does not use every data qubit at every timestep.
It is prevented from doing so by the full use of every ancilla qubit at every timestep.
With this in mind, we consider an alternative parity check circuit in which the weight-eight octagon checks are measured using a Bell state ancilla, i.e. $\ket{+}$ is replaced by $\frac{1}{\sqrt{2}}(\ket{00}+\ket{11})$. One can decrease the length of the \cnot schedule to four, with two data qubits interacting with each octagon ancilla state at every timestep. The resulting circuit is shown in Figure \ref{fig:4_step_sched}.
\begin{figure}
\centering
\includegraphics[width=\textwidth]{four_step_figure.pdf}
\vspace*{10pt}\fcaption{Four-step data-ancilla gate schedule for syndrome extraction in the \name.
Pairs of octagon ancillas are prepared in Bell states prior to interaction with the data qubits, and their measurement outcomes ($\pm 1$) are multiplied to obtain the syndrome.
Note that the square $X$ and $Z$ checks are measured simultaneously.
This syndrome extraction requires two ancillas per square and octagon, for a total of $2d^2$ ancillas, identical to the toric code. }
\label{fig:4_step_sched}
\end{figure} 
Using this \cnot schedule, the threshold of the \name is appreciably increased, as seen in Figure \ref{fig:full_monty}.
\begin{figure}
\centering

\vspace*{10pt}\fcaption{Logical error probability of the \name with the four-step schedule in Figure~\ref{fig:4_step_sched}.
The threshold error probability has increased to $\sim 0.41\%$.
Error bars are $95\%$ confidence intervals.}
\label{fig:full_monty}
\end{figure}

Though this threshold is still lower than that of the surface code, it far exceeds previous thresholds for color codes.
In the following section, we discuss the implications of this moderate threshold value for other error-correcting codes and for hardware architectures with clustered layouts.

\section{Discussion \& Future Work}
\label{sec:discussion}
To obtain more information about when the \name{} could be preferred over the toric code, we examine the consequences of tripling the error rates of \cnot gates associated with octagons. In other words, the depolarizing noise rate on all gates is $p$ except for the CNOT gates used to perform the octagon parity checks where it is $3p$. As we see in Figure \ref{fig:oct_factor_3}, the threshold $p_c$ of the \name{} becomes $\sim 0.21\%$. This implies that one can tolerate a `long-range' CNOT error rate of $\sim 0.63\%$, {\em above} the toric threshold, as long as the short-range CNOT error rate (and all other operations) is a factor of 3 lower.
As a result we expect that an error rate difference appreciably higher than a factor of 3 will favour the use of a code like the \name.
Greater precision in this estimate depends on altering the minimum-weight perfect matching algorithm to use edge weights derived from error propagation through the syndrome extraction circuit \cite{PhysRevA.83.020302}.
\begin{figure}
\centering

\fcaption{Logical error probability of the \name{} with the four-step schedule in Figure \ref{fig:4_step_sched}, with a tripled error rate for all \cnot gates supported on an ancilla inside an octagon. 
This represents a possible error rate discrepancy for long-range gates. 
Error bars are $95\%$ confidence intervals.}
\label{fig:oct_factor_3}
\end{figure}

An open question is whether our results can be extended to show that the full square-octagon color code can have a threshold close to that of the toric code when correcting circuit-based noise.
This would imply that one can both get the advantage of transversal gates and a good threshold.
However we suspect that the good performance may relate to using the gauge qubit degree of freedom in the decoder and in designing the parity check circuits.
Further optimization of the color code decoder in \cite{MultiMWPMDecoder, stephens:threshold}, possibly using insights of this paper, would be able to shed light on this question.

Let us add a few comments on a potential hardware advantage of the \cfour{} toric code for superconducting transmon qubits.
With these qubits, several types of two-qubit gates have been realized.
The two-qubit \cz gate which uses a direct capacitive coupling between two transmon qubits has been reported to have a low error rate of $0.6\%$ and short duration $< 50$ ns \cite{barends+:jos_surf}.
The range of this gate is determined by the size of the transmon qubits, approximately 200 $\mu$m in \cite{kelly+:repetition} and the gate requires flux-tunable qubits (hence space for flux lines).
A surface code implementation which only uses this \cz gate seems to hard to achieve in a purely 2D fashion, as the qubits will be closely packed together: see for example Figure 1 in \cite{kelly+:repetition} in which one has a 1D array of capacitively-coupled qubits which can be used to implement the parity checks of the repetition code but not those of the 2D surface code. 
The alternative is to use a bus-resonator-mediated two-qubit gate.
Such a resonator-mediated gate can be activated by only microwave control such as in the cross-resonance gate used by the IBM group, see \cite{corcoles:detecting, chow+:4qubit} with a currently reported error rate of $1.4\%$ \cite{GCS:overview}.
The resonator-mediated coupling can also be used to enact an iSWAP gate where flux-tunable qubits are brought into resonance, or the qubit state can be explicitly put on the resonator bus as in the two-qubit gate employed in \cite{riste:detecting} (a variety of other two-qubit gates are listed in \cite{GCS:overview}).
A resonator-mediated gate is the basic building block in the surface code layout for transmon qubits as described in \cite{divincenzo_arch}.
We may thus speculate that it is of interest to use short-range, short-duration, high-fidelity \cz gates between directly coupled qubits, for measuring, say, the square $XXXX$ and $ZZZZ$ checks.
To measure the octagonal checks between the four-qubit cluster one can consider using longer-range bus-resonator mediated gates (a bus resonator at $8$GHz has physical length of $\sim 1$ cm which can be compactly wrapped-up) allowing for more flexibility.
The octagonal checks could also be measured as {\em direct} parity measurements (see references to direct parity check measurements in \cite{Terhal:RMP}).
As long as long-range operations have error rates appreciably higher than those at short range, we suspect that codes such as the \name{} will be applicable. 

\nonumsection{Acknowledgements}
\noindent
The authors thank Nikolas Breuckmann and Kasper Duivenvoorden for useful discussions, and Tobias H\"{o}lzer for his early work on this project.
This work is supported by the European Research Council (EQEC, ERC Consolidator Grant No: 682726).
B. Criger acknowledges financial support through the program SCALEQIT, and computing support from the RWTH ITC.

\nonumsection{Comments}
\noindent
Numerical simulations were performed using \texttt{py-qcode}, which can be found at \\ \texttt{github.com/bcriger/py-qcode}.

\section*{References}
\bibliographystyle{hunsrt}
\bibliography{Enhanced_Thresholds}

\begin{appendices}
\section{Concatenation Construction of the 6.6.6 Color Code}
\label{app:642}
We place the physical qubits of the $\llbracket 6,4,2\rrbracket$ code on the vertices of red plaquettes of a hexagonal lattice, expressing the checks and logical operators as in Figure \ref{fig:422} above:
\begin{figure}[H]
\centering
\begin{tikzpicture}
\foreach \y in {0, 1}{
    \foreach \x in {0, 1, ..., 6}{
        \node (hex_\x_\y) [regular polygon, regular polygon sides=6, draw, line width=2pt, scale=4, line join=round, minimum size=0.3cm, outer sep=0pt] at (2.1*\x,2*\y) {};
    }
}
\foreach \x/\xlab/\zlab in {0/$S_X$/$S_Z$, 1/$\overline{X}_1$/$\overline{Z}_1$,
                           2/$\overline{X}_2$/$\overline{Z}_2$, 3/$\overline{X}_3$/$\overline{Z}_3$, 4/$\overline{X}_4$/$\overline{Z}_4$, 5/$\overline{X}_2 \overline{X}_3 S_X$/$\overline{Z}_2 \overline{Z}_3 S_Z$, 6/$\overline{X}_1 \overline{X}_4 S_X$/$\overline{Z}_1 \overline{Z}_4 S_Z$}{
    \node[above=6pt] at(hex_\x_1.north) {\xlab};
    \node[below=6pt] at(hex_\x_0.south) {\zlab};
}

\node[regular polygon, regular polygon sides=6, draw, line width=14pt, scale=2, line join=round] at (0,0) {};
\node[regular polygon, regular polygon sides=6, draw, line width=12pt, scale=2, line join=round, white] at (0,0){};
\node(z_stab)[regular polygon, regular polygon sides=6, draw=none, line width=10pt, scale=2, line join=round, white, inner sep=0pt] at (0,0){};
\foreach \n [count=\nu from 0, remember=\n as \lastn, evaluate={\nu+\lastn}] in {1,2,...,6} 
\node at(z_stab.corner \n){$Z$};

\node[regular polygon, regular polygon sides=6, draw, line width=14pt, scale=2, line join=round] at (0,2) {};
\node[regular polygon, regular polygon sides=6, draw, line width=12pt, scale=2, line join=round, white] at (0,2){};
\node(x_stab)[regular polygon, regular polygon sides=6, draw=none, line width=10pt, scale=2, line join=round, white, inner sep=0pt] at (0,2){};
\foreach \n [count=\nu from 0, remember=\n as \lastn, evaluate={\nu+\lastn}] in {1,2,...,6} 
\node at(x_stab.corner \n){$X$};
\foreach \x/\s/\e in {1/1/2, 2/6/1, 3/4/5, 4/3/4, 5/2/3, 6/5/6}{
\path[draw=black, line width = 14pt, line cap=round] (hex_\x_0.corner \s) -- (hex_\x_0.corner \e);
\path[draw=white, line width = 12pt, line cap=round] (hex_\x_0.corner \s) node {$Z$} -- (hex_\x_0.corner \e) node {$Z$};
}
\foreach \x/\s/\e in {1/6/1, 2/1/2, 3/3/4, 4/4/5, 5/5/6, 6/2/3}{
\path[draw=black, line width = 14pt, line cap=round] (hex_\x_1.corner \s) -- (hex_\x_1.corner \e);
\path[draw=white, line width = 12pt, line cap=round] (hex_\x_1.corner \s) node {$X$} -- (hex_\x_1.corner \e) node {$X$};
}
\end{tikzpicture}
\vspace*{10pt}\fcaption{Stabilizer checks and logical operators for the $\llbracket 6,\,4,\,2 \rrbracket$ code, placed on the vertices of a hexagon. 
The operators $XX$ and $ZZ$, when placed on an edge, are logical operators, though some are products of two generators.
}
\label{fig:642}
\end{figure}
The color code checks associated with green and blue plaquettes can be written as tensor products of these edge operators:
\begin{figure}[H]
\centering
\begin{tikzpicture}
\foreach \x/\col/\y in {0/gn/0, 1/bl/-0.65cm, 2/gn/0, 3/bl/-0.65cm}{ 
\node (hex_\x) [regular polygon, regular polygon sides=6, draw, fill=\col, line width=2pt, scale=4, line join=round, minimum size=0.3cm, outer sep=0pt] at (3.5*\x,\y) {};
\node[regular polygon, regular polygon sides=6, draw, line width=14pt, scale=2, line join=round] at (3.5*\x,\y) {};
\node[regular polygon, regular polygon sides=6, draw, line width=12pt, scale=2, line join=round, white] at (3.5*\x,\y){};
\node(stab_\x)[regular polygon, regular polygon sides=6, draw=none, line width=10pt, scale=2, line join=round, white, inner sep=0pt] at (3.5*\x,\y){};
}
\foreach \n [count=\nu from 0, remember=\n as \lastn, evaluate={\nu+\lastn}] in {1,2,...,6}{
\node at(stab_0.corner \n){$Z$};
\node at(stab_1.corner \n){$Z$};
\node at(stab_2.corner \n){$X$};
\node at(stab_3.corner \n){$X$};
}

\foreach \x in {0, 2}{
\node[anchor=side 1] (hex_\x_bot) [regular polygon, regular polygon sides=6, draw, fill=rd, line width=2pt, scale=4, line join=round, minimum size=0.3cm, outer sep=0pt] at (hex_\x.side 4) {};
\node[anchor=side 5] (hex_\x_upl) [regular polygon, regular polygon sides=6, draw, fill=rd, line width=2pt, scale=4, line join=round, minimum size=0.3cm, outer sep=0pt] at (hex_\x.side 2) {};
\node[anchor=side 3] (hex_\x_upr) [regular polygon, regular polygon sides=6, draw, fill=rd, line width=2pt, scale=4, line join=round, minimum size=0.3cm, outer sep=0pt] at (hex_\x.side 6) {};
}
\foreach \x in {1, 3}{
\node[anchor=side 4] (hex_\x_top) [regular polygon, regular polygon sides=6, draw, fill=rd, line width=2pt, scale=4, line join=round, minimum size=0.3cm, outer sep=0pt] at (hex_\x.side 1) {};
\node[anchor=side 2] (hex_\x_lor) [regular polygon, regular polygon sides=6, draw, fill=rd, line width=2pt, scale=4, line join=round, minimum size=0.3cm, outer sep=0pt] at (hex_\x.side 5) {};
\node[anchor=side 6] (hex_\x_lol) [regular polygon, regular polygon sides=6, draw, fill=rd, line width=2pt, scale=4, line join=round, minimum size=0.3cm, outer sep=0pt] at (hex_\x.side 3) {};
}
\node[white] at (hex_0_bot.center) {$\overline{Z}_1$};
\node[white] at (hex_0_upl.center) {$\overline{Z}_1\overline{Z}_4$};
\node[white] at (hex_0_upr.center) {$\overline{Z}_4$};
\node[white] at (hex_2_bot.center) {$\overline{X}_2$};
\node[white] at (hex_2_upl.center) {$\overline{X}_2\overline{X}_3$};
\node[white] at (hex_2_upr.center) {$\overline{X}_3$};
\node[white] at (hex_1_top.center) {$\overline{Z}_3$};
\node[white] at (hex_1_lor.center) {$\overline{Z}_2\overline{Z}_3$};
\node[white] at (hex_1_lol.center) {$\overline{Z}_2$};
\node[white] at (hex_3_top.center) {$\overline{X}_4$};
\node[white] at (hex_3_lor.center) {$\overline{X}_1\overline{X}_4$};
\node[white] at (hex_3_lol.center) {$\overline{X}_1$};
\end{tikzpicture}
\vspace*{10pt}\fcaption{Green and blue checks of the 6.6.6 color code, expressed using edge operators of the $\llbracket 6,\,4,\,2 \rrbracket$ code given in Figure \ref{fig:642}.
When expressed in this fashion, each of these checks has weight four.}
\label{fig:neighbour_hexes}
\end{figure}
The green and blue checks of the 6.6.6 color code divide into disjoint pairs, those supported on logical qubits $1$ and $4$ for every red hexagon, and those supported on logical qubits $2$ and 3.
Placing two physical qubits on each red hexagon, we can see that the weight-four checks on this lattice create a toric code:
\begin{figure}[H]
\centering
\begin{tikzpicture} [hexa/.style= {shape=regular polygon,regular polygon sides=6,minimum size=2.5cm, draw=black!30, inner sep=0,anchor=center,line width=2pt}, scale=2.5]
\begin{pgfonlayer}{fg}
\clip[draw=none] ({(4) *0.5* (1 + sin(30))},{(2) *0.5* cos(30)}) rectangle ({(8) *0.5* (1 + sin(30))},{(-4) *0.5* cos(30)});
\end{pgfonlayer}
\clip[draw=none] ({(4) *0.5* (1 + sin(30))},{(2) *0.5* cos(30)}) rectangle ({(8) *0.5* (1 + sin(30))},{(-4) *0.5* cos(30)});

\foreach \j in {0,...,6}{
    \foreach \i in {0,...,6}{%
        \node[hexa] (h\i;\j) at ({(\i + \j) *0.5* (1 + sin(30))},{(\i - \j) *0.5* cos(30)}) {};
    }
}
\begin{pgfonlayer}{fg}
\foreach \i/\j in {0/1, 0/4, 1/2, 1/5, 2/0, 2/3, 2/6, 3/1, 3/4, 4/2, 4/5, 5/0, 5/3, 5/6, 6/1, 6/4}{
	\node [fill=rd, draw=none, shape=regular polygon,regular polygon sides=6,minimum size=2.5cm, inner sep=0, opacity=0.5] at (h\i;\j.center) {};  
    \node (q\i;\j;2) [above=2, circle, draw, fill=white] at (h\i;\j.center) {2};
    \node (q\i;\j;3) [below=2, circle, draw, fill=white] at (h\i;\j.center) {3};
}
\end{pgfonlayer}
\foreach \i/\j in {0/1, 0/4, 1/2, 1/5, 2/3, 3/1, 3/4, 4/2}{
    \pgfmathtruncatemacro{\ipone}{\i + 1}
    \pgfmathtruncatemacro{\iptwo}{\i + 2}
    \pgfmathtruncatemacro{\jpone}{\j + 1}
    \pgfmathtruncatemacro{\jmone}{\j - 1}
    \filldraw[line width=2pt, fill=bl, fill opacity=0.5] (q\i;\j;2.center) -- (q\iptwo;\jmone;3.center) -- (q\ipone;\jpone;2.center) -- (q\ipone;\jpone;3.center) -- cycle;    
}
\foreach \i/\j in {2/0, 3/1, 4/2, 1/2, 2/3, 3/4}{
    \pgfmathtruncatemacro{\ipone}{\i + 1}
    \pgfmathtruncatemacro{\imone}{\i - 1}
    \pgfmathtruncatemacro{\jpone}{\j + 1}
    \pgfmathtruncatemacro{\jptwo}{\j + 2}
    \filldraw[line width=2pt, fill=gn, fill opacity=0.5] (q\i;\j;2.center) -- (q\i;\j;3.center) -- (q\imone;\jptwo;2.center) -- (q\ipone;\jpone;3.center) -- cycle;    
}
\filldraw[line width=2pt, fill=gn, fill opacity=0.5] (q0;4;2.center) -- (q0;4;3.center) -- (q1;5;3.center) -- cycle;
\filldraw[line width=2pt, fill=gn, fill opacity=0.5] (q1;5;2.center) -- (q1;5;3.center) -- (q2;6;3.center) -- cycle;
\end{tikzpicture}
\vspace*{10pt}\fcaption{Toric code stabilizers supported on a 6.6.6 lattice with two physical qubits per red hexagon, qubits $2$ and $3$ from Figure \ref{fig:neighbour_hexes}.}
\label{fig:transparent_hexes}
\end{figure}
\end{appendices}

\end{document}